\pdfoutput=1
\documentclass[aps,pre, twocolumn, groupedaddress]{revtex4-1}

\usepackage{lipsum}
\usepackage{mathtools}
\usepackage{graphicx}
\usepackage{dcolumn}
\usepackage{amsmath}    
\usepackage{amssymb}
\usepackage{bm}
\usepackage{hyperref}
\usepackage{latexsym}
\usepackage{verbatim}
\usepackage[normalem]{ulem}
\usepackage{color}
\usepackage[caption=false]{subfig}
\def\beq{\begin{equation}}
\def\eeq{\end{equation}}

\def\beq{\begin{equation}}                          
\def\eeq{\end{equation}}                          
\def\bea{\begin{eqnarray}}                          
\def\eea{\end{eqnarray}}

\DeclareRobustCommand{\uvec}[1]{{%
  \ifcsname uvec#1\endcsname
     \csname uvec#1\endcsname
   \else
    \bm{\hat{\mathbf{#1}}}%
   \fi
}}
\preprint{}
\begin{document}
\preprint{}
\title{Ordering kinetics and steady state of  self-propelled particles with random-bond disorder}
\author{Jay Prakash Singh$^{1}$}
\email{jayps.rs.phy16@itbhu.ac.in}
\author{Sudipta Pattanayak$^{2}$}
\email{sudipta.pattanayak@bose.res.in}
\author{Shradha Mishra$^{1}$}
\email{smishra.phy@iitbhu.ac.in}
\affiliation{$^{1}$Department of Physics, Indian Institute of Technology (BHU), Varanasi, India 221005}
\affiliation{$^{2}$S. N. Bose National Centre for Basic Sciences, J D Block, Sector III, Salt Lake City, Kolkata 700106}
\date{\today}
\begin{abstract}
In this study, we introduce a minimal model for a collection of polar 
	self-propelled particles (SPPs) on a two-dimensional
	substrate where each particle has a different ability to interact with its neighbours. The SPPs 
	 interact through a short-range alignment interaction and interaction strength of each
	 particle is obtained from a uniform distribution. Moreover, 
	 the volume exclusion among the SPPs is taken care of
	 by introducing a repulsive interaction among them.
	We characterise the ordered steady state and kinetics of the system for different strengths of the disorder.
	We find that the presence of the disorder does not destroy the usual 
	long-range ordering in the system.
	To our surprise, we note that the density clustering is enhanced in the presence of the disorder.
	Moreover, the disorder leads  to the formation of a random network of different 
	interaction strengths, which makes the alignment weaker and 
	it results in the slower dynamics. 
	Hence, the disorder leads to more cohesion among the particles.  
	Furthermore, we note that the kinetics of the ordered 
	state remains unaffected in the presence of the disorder.
	Size of orientationally ordered domains 
	and density clusters grow with time 
	with dynamic growth exponents $z_{o} \sim 2$ and 
	$z_{\rho} \sim 4$, respectively. 
\end{abstract}
\maketitle
\section{Introduction \label{Introduction}}
Collective behaviour of a large number of 
self-propelled particles (SPPs) or ``flocking'' is ubiquitous  in nature. Examples of such
systems range from a few micrometres, e.g., actin and tubulin filaments,
molecular motors, epithelial cells \cite{Nedelec1997, Yokota1986, Garcia}, unicellular organisms such as amoebae and
bacteria \cite{Bonner1998}, to several metres, e.g., birds flock \cite{Chen2019}, fish school \cite{Parrish1997}
and human crowd \cite{Helbing2000} \textit{etc}. Interestingly, these systems show a collective motion 
on a scale much larger than each individual even in two dimensions, hence, long-range ordering (LRO) 
is observed. A minimal model was introduced by Vicsek \textit{et al.} 
to understand the basic features of the collective behavior 
of self-propelled polar particles or ``polar flock'' \cite{VicsekT}.
 In the last three decades, many variants of the Vicsek model
are studied to understand various features of different model systems \cite{Chateprl2004, Chatepre2008, SudiptaJPCOM, Ihlepre2014}. \\
 In these studies, authors mainly consider a collection of SPPs in a homogeneous system or medium.
Recently, there is a growing interest to understand the effects and advantages of different kinds of inhomogeneities
which are omnipresent in nature.
Many studies show that the inhomogeneity can destroy the LRO present
in a disorder-free system \cite{Morin2017, Chepizhko2013, Yllanes2017, Quint2015, Sandor2017, Reichhardt2017, Rakesh2018, 
Toner2018E, Toner2018L} whereas a few studies discuss 
special kinds of inhomogeneities which can enhance the ordering of a system \cite{RDas2020, SudiptaIS}.
Therefore, the inhomogeneity can be useful for many practical 
applications, e.g., crowd control and faster evacuation etc.\cite{Lin2018, Dorso2011, 
ZuriguelJSM, Zuriguel2016, Zuriguel2011}. \\
 In the Vicsek model, each individual interacts through a short-range
alignment interaction and the strength of the interaction
is the same for all  the particles.
But, in natural systems, each particle can have a different ability to influence its neighbors.
However, scientists have not paid much attention to understand
the effects of different interaction strengths in a polar flock. 
In a recent study, Bialek {\em et al.} show that pairwise inhomogeneous interactions between particles are 
sufficient to correctly predict the propagation of order throughout the entire flock \cite{WilliamPNAS}. \\
In this work, we introduce a collection of polar SPPs with the random-bond disorder, and
the particles interact through a short-range alignment interaction. 
Moreover, the volume exclusion among the 
particles is taken care of by introducing a repulsive interaction among them 
\cite{Dgayer,Petitjean,Caprini}.
The strength of interaction for each particle
is obtained from a uniform distribution between $[1-\epsilon/2: 1+\epsilon/2]$, 
where $\epsilon$ is the  
strength of random-bond disorder. 
For $\epsilon=0$, the model represents a disorder-free 
polar flock with uniform interaction strength for 
all the particles or the Vicsek-like model \cite{VicsekT}.
In this study, our focus is to understand the
effects of the random-bond disorder on
 the true long-range ordered state in a disorder-free system \cite{VicsekT, TonerTu1998}.
Also, we have characterised the effects of the random-bond disorder
on the ordering kinetics of a polar flock. \\
We note that the presence of the disorder does not destroy 
the LRO present in a disorder-free system.
However, the
disorder affects the density clustering and results in more {\it cohesive} flocking.
Furthermore, we also studied
the ordering kinetics of the orientation and the density fields.
When the system is quenched from an isotropic to an ordered  steady state,
both the orientation and the density fields coarsen with time. The 
size of the ordered orientation domains
grows with time with an effective growth exponent $z_o \sim 2$ 
(same as for non-conserved model A \cite{Bray1994, Puri2009}. Also,
the size of the high-density domains grow with time with an exponent 
$z_{\rho} \sim 4$, similar to as found for a conserved field
in active systems \cite{Cates2014, SudiptamodelB}. \\

The rest of the paper is organised as follows. In Sec.\ref{model}, we discuss
the model and simulation details. In Sec.\ref{results}, the results from the 
numerical simulations are discussed.
In Sec.\ref{Discussion}, we conclude the paper with
a summary and discussion of the results.
Appendix \ref{App:AppendixA} includes the details of linearised hydrodynamics to
calculate the local density fluctuations in the system.
\section{Model \label{model}}
\begin{figure}[t]
\centering
\includegraphics[width=1.0\linewidth]{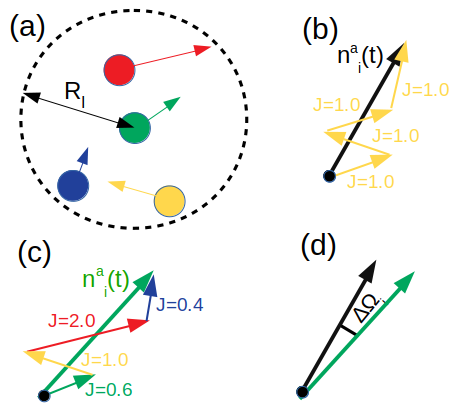}
	\caption{(color online) (a) Cartoon picture
        of the model. The dashed circle of radius $R_{I}$
        represents the interaction radius of
        the green tagged particle of radius $R$ (at the centre).
        The circles of various colors of radius $R$ indicate the neighbors of
        the tagged particle. The arrows of different lengths represent
        the interaction strength $J'$s of the respective particle.
        (b,c) The cartoon picture of the resultant
        direction of the tagged particle due to the alignment
        interaction with its neighbors for the uniform strength (clean polar flock) and
        the varying (RBDPF) interaction strength model, respectively. Black and green
        arrow represent the resultant
        directions of the tagged particle in (b) and (c), respectively.
        (d) The relative difference in the resultant direction $\Delta\Omega_i$
        of the tagged particle for the clean and
        the RBDPF.}
\label{fig:fig1}
\end{figure}

We consider a collection of $N$ polar self-propelled particles (SPPs) moving on a two-dimensional
substrate. SPPs  interact through a short-range alignment interaction within  interaction
radius $R_I$ \cite{VicsekT,Chateprl2004,Chatepre2008}. 
Moreover, the strength of interaction of each SPP is {\em different} unlike the Vicsek model 
of uniform interaction strength \cite{VicsekT}. 
Furthermore, the volume exclusion among the particles is introduced through a soft repulsive 
binary force ${\bf f}_{ij}$,
to avoid the clustering of particles to a single point for low noise or strong alignment \cite{Lucas}. 
Each SPP is defined by its position 
${\bf r}_i$ and orientation $\theta_{i}$, and it moves
along its direction vector ${\bf n}_{i}(t) = (\cos(\theta_i(t)), \sin(\theta_i(t)))$ with a fixed speed $v_0$. 
The two update equations for the position ${\bf r}_i(t)$ and 
the direction vector ${\bf n}_{i}(t)$
are given by, 
\begin{equation}
	{\bf {r}}_{i} (t + \Delta{t}) = {\bf {r}}_i(t) + {v_{0}}{\bf n}_{i}(t){\Delta{t}}  
\label{eq1}
\end{equation}
\begin{equation}
	{\bf n}_{i}(t+\Delta{t})=\frac {{}\sum _{{j\in R_{I}}}J_{j}{\bf n}_{j}(t)-\beta\sum_{j\in R}{{\bf f}_{ij}}+\eta N_{i}(t){\bf {\bf{{\xi}}}}_{i}(t)}{w_i{(t)}}
\label{eq2}
\end{equation}
and soft repulsion force
$	{\bf f}_{ij}=\bigg(\exp\big{[}1-({\frac{r_{ij}}{R})}^{\gamma}\big{]}-1\bigg){{\bf e}_{ij}}$,
where ${\bf f}_{ij} \neq 0$   if  ${r_{ij}} < R$, and ${\bf f}_{ij} = 0$ if $r_{ij} \geq R$, where $R = R_{I}/10$ is the typical size 
of the particles. 
$r_{ij} =\mid{\bf r}_{j}-{\bf r}_{i}\mid$, ${\bf e}_{ij} = \frac{{\bf r}_{ij}}{r_{ij}}$ and 
 the exponent $\gamma = 0.25$ is kept fixed such that the range of the repulsive force is smaller than the $R_I$. \\
 Eq.(\eqref{eq1}) represents the motion of the  particle due to its self-propelled nature along the direction vector ${\bf n}_i(t)$ with a fixed speed $v_0$. $\Delta t =1.0$ is the unit time step. 
The first term on the right hand side in Eq.(\eqref{eq2}) represents the short-range alignment interaction of the $i^{th}$ particle with its 
neighbors within the interaction radius ($R_{I}=1.0$),  
and $J_{j}$ is the  interaction strength of the $j^{th}$ neighbor.
The probability distribution of the interaction strength 
$J$, $P(J)$, is obtained from a uniform distribution of
range $[1-\frac{\epsilon}{2}:1+\frac{\epsilon}{2}]$ \cite{MKumar}, 
where $\epsilon$ measures the degree of disorder. 
$\epsilon=0$ corresponds to the uniform
interaction strength $(J_{i}=1$ for all the particles) 
like the Vicsek model \cite{VicsekT} whereas 
$\epsilon=2$ corresponds to the maximum 
disorder in the system. 
The second term indicates the 
soft-repulsive force due to the finite size of the 
particles. The strength of the force $\beta$ is kept fixed to $0.01$.
Furthermore, the third term in the 
 Eq.(\eqref{eq2}) denotes the vector noise 
which measures the error made by the particle while
following its neighbors. 
$\xi_i(t)$ is a random unit vector and 
$N_{i}(t)$ denotes the number of neighbors within 
the interaction radius of the $i^{th}$ particle at time $t$.
$\eta$ represents the strength of the noise and it can vary from
$0$ to $1$. $w_{i}(t)$ is the normalisation factor, which 
reduces the R. H. S. of the Eq.(\eqref{eq2}) to a unit vector. \\
The cartoon picture of the  model is shown
in Fig.\ref{fig:fig1} (a). The resultant
direction vector 
${\bf n}^{a}_{i}(t)$ 
of the 
$i^{th}$ particle (due to alignment interaction with its neighbors) 
for the disorder-free ($\epsilon=0$) and 
the maximum disorder ($\epsilon=2.0$) 
system are shown in Fig.\ref{fig:fig1}(b) and (c), respectively. 
In Fig.\ref{fig:fig1}(d), $\Delta \Omega_{i}$ represents 
the difference in the resultant vectors shown in Fig. \ref{fig:fig1}(b) and (c). For a disorder 
system the resultant direction vector is closer to the particle's original direction, which 
is due  to the weaker alignment in the presence of disorder.
In analogy with the  equilibrium random-bond $XY$(RBXY)-model \cite{Zh, MKumar},
we name our  model as random-bond disorder in polar flock (RBDPF).
However, for $\epsilon=0$, the model reduces to a disorder-free or {\em clean} polar flock.
We numerically update the  Eqs.(\eqref{eq1}) and (\eqref{eq2}) 
for all SPPs sequentially. One simulation step 
is counted after the update of Eqs.(\eqref{eq1}) and (\eqref{eq2}) once for
all the particles. Periodic boundary conditions (PBC) 
are used for a system  of size 
$L \times L$, and $L$ is varied from
$50$ to $512$ ($N$ from $2500$ to $262144$). 
The number density of the system is 
defined as $\rho _0 = \frac{N}{L \times L}$.
Most of the results are obtained for  density  $\rho _0 = 1.0$ and 
some results  are calculated for $\rho_0 = 0.5$ and $2.0$. The self-propulsion speed is fixed at $v_0 = 0.5$. 
The noise strength $\eta$ is fixed at $\eta = 0.2$, 
such that the steady state is 
an ordered state and the system is away from the 
 order-disorder phase transition \cite{Chatepre2008}. 
The effect of random bond disorder on the system near order-disorder phase transition will be our future study \cite{jpsingh}.
We study the properties of steady state as well as 
the ordering kinetics of the orientation and density fields for different strengths of the 
disorder $\epsilon$. 
We consider time up to $10^4$ to study the ordering kinetics and steady state 
results are obtained from time up to $10^6$ and $20$ independent realizations 
are used for the better statistics of the numerical results.
\begin{figure}[t]
\centering
\includegraphics[width=1.0\linewidth]{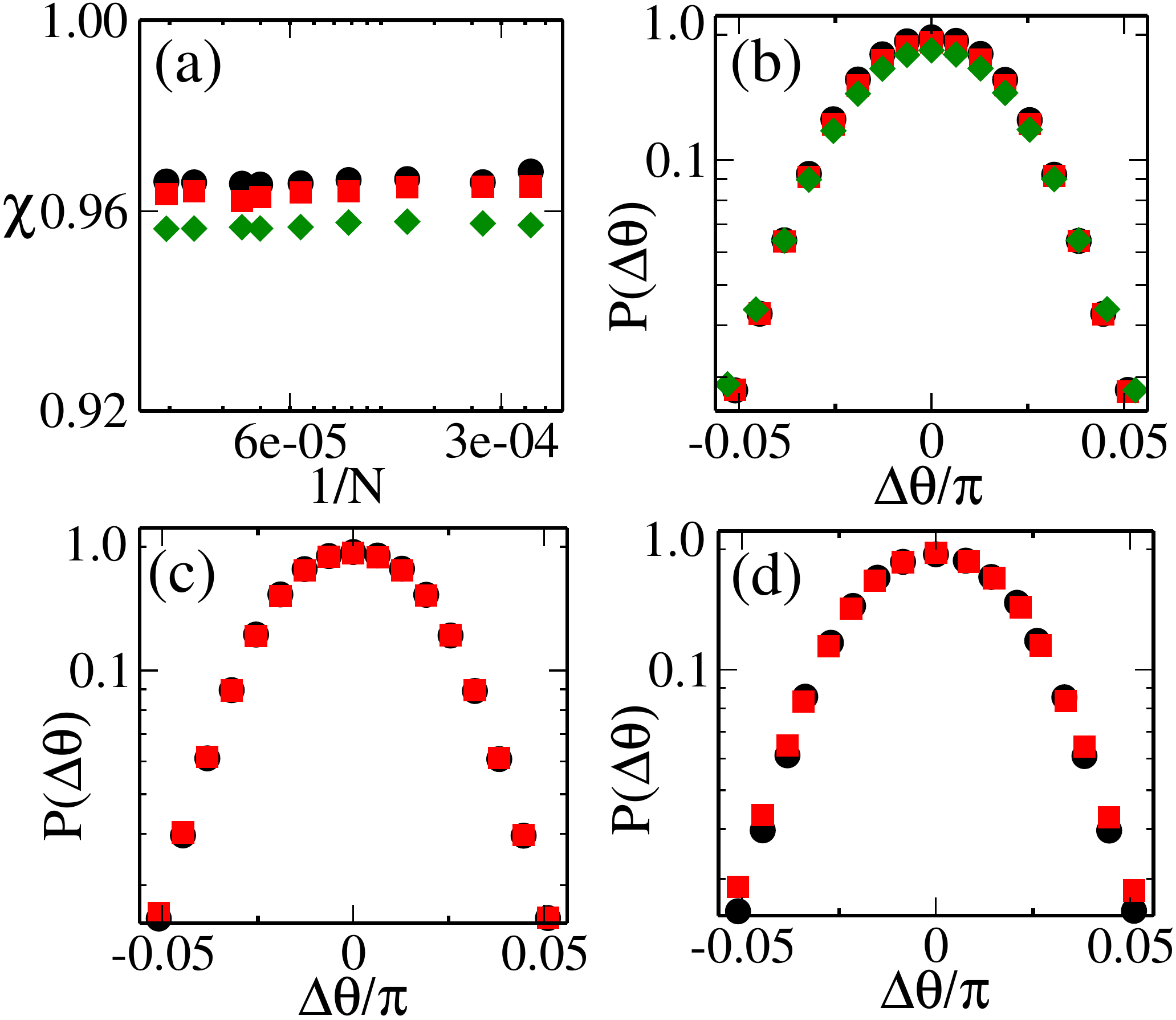}
\caption{(color online) (a) Plot of the global orientation
        order parameter $\chi$ vs. $1/N$
        for different $\epsilon$ in semi$-\log$ X scale.
        (b) Probability distribution function of the mean
        orientation fluctuation $P(\Delta\theta)$ vs. $\frac{\Delta\theta}{\pi}$ 
	for different $\epsilon$ in semi$-\log$ Y scale. $N=62500$.
        The filled black circles, red squares,
        green diamonds represent data for $\epsilon=0.0, 1.0$,
        and $2.0$, respectively.
        (c,d) Plots of $P(\Delta\theta)$ vs. $\frac{\Delta\theta}{\pi}$ for different
        system sizes for $\epsilon=0.0$ and $\epsilon=2.0$  in semi$-\log$ Y scale, respectively.
        The filled black circles and red squares denote
        $N = 40000$ and $62500$, respectively.}
\label{fig:fig2}
\end{figure} 
\section{Results\label{results}}
\subsection{Steady-state behaviour\label{Steady-state behaviour}}
In uniform-interaction strength models or Vicsek-like models
\cite{VicsekT, Chateprl2004, Chatepre2008}, the ordered state exhibits a true long-range
order in two dimensions.
In general, the orientation ordering in the system 
is characterised by the global 
orientation order parameter, which is defined as,  
 $\chi(t)=\frac{1}{N}|{{\sum}^N_{i=1}{{\bf n}_i}}(t)|$.
 $\chi(t)$ is very small and it is of the order $\frac{1}{\sqrt{N}}$ for the disordered state and it is
 close to unity in the ordered state. 
 The variation of the mean value of $\chi(t)$, $\chi$, {\em vs.} $1/N$ for different $\epsilon$ is shown
 in Fig.\ref{fig:fig2}(a), where ``mean'' is obtained from the value of $\chi(t)$ in the steady state and it is averaged over
  $20$ independent realisations.
 We note that $\chi$ is independent of system size for different strengths
 ($\epsilon$) of the disorder. 
 However, the magnitude of $\chi$ shows a small variation 
 on increasing the strength $\epsilon$ of the disorder. 
 Furthermore, the probability distribution function (PDF) of fluctuation from the mean
 orientation of the particles $P(\Delta\theta)$ is shown for 
 different values of $\epsilon$ in 
  Fig.\ref{fig:fig2}(b), where $\Delta\theta=\theta_i-\overline{\theta}$ where, $\theta_i$ is the orientation of $i^{th}$ particle and $\overline{\theta}$ is mean orientation of the flock.
 The peak of the PDF decreases with the increasing disorder strength 
 $\epsilon$. Moreover, the change is small but it is 
 consistent with increasing $\epsilon$. 
 To confirm the long-range ordering, 
 we plot $P(\Delta\theta)$ for different system sizes
 for $\epsilon=0$ and $2$, 
 in Fig.\ref{fig:fig2}(c) and (d), respectively. 
 $P(\Delta\theta)$ distribution for different system sizes 
 overlaps on each other for a particular $\epsilon$. 
Therefore, the magnitude of the global ordering shows a small decay 
with increasing $\epsilon$ but the ordered steady state 
remains long range for all $\epsilon$ of RBDPF.
\begin{figure}[t]
\centering
\includegraphics[width=1.0\linewidth]{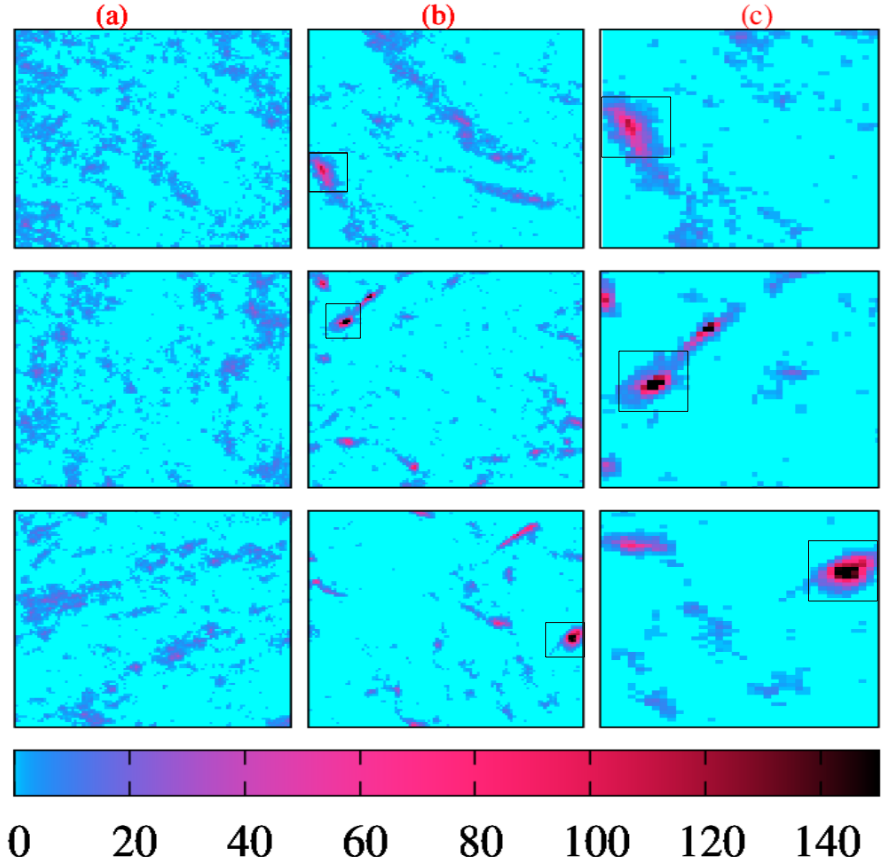}
	\caption{(color online) Horizontal panel: top to bottom panels are real space snapshots of the local number density of the SPPs for different $\epsilon$ at different times t. The topmost panel is for $\epsilon = 0$, middle one is for  $\epsilon= 1$ and bottom is for  $\epsilon= 2$. Vertical panels:  from left to right, (a) to (c), are real snapshots of the local number density of the SPPs at different time t for each $\epsilon$. Leftmost panel (a) is for t = 5,000, the middle one (b) is for t = 45,000 and (c) represent zoomed snapshots of (b) at time t=45000. The square boxes in (c) represent the zoomed version of the square boxes of (b). N = 10,000. The color bar represents the local number density of the particles. .}
\label{fig:fig3}
\end{figure}
\subsubsection*{Behaviour of the flock state}
\begin{figure}[t]
\centering
\includegraphics[width=1.0\linewidth]{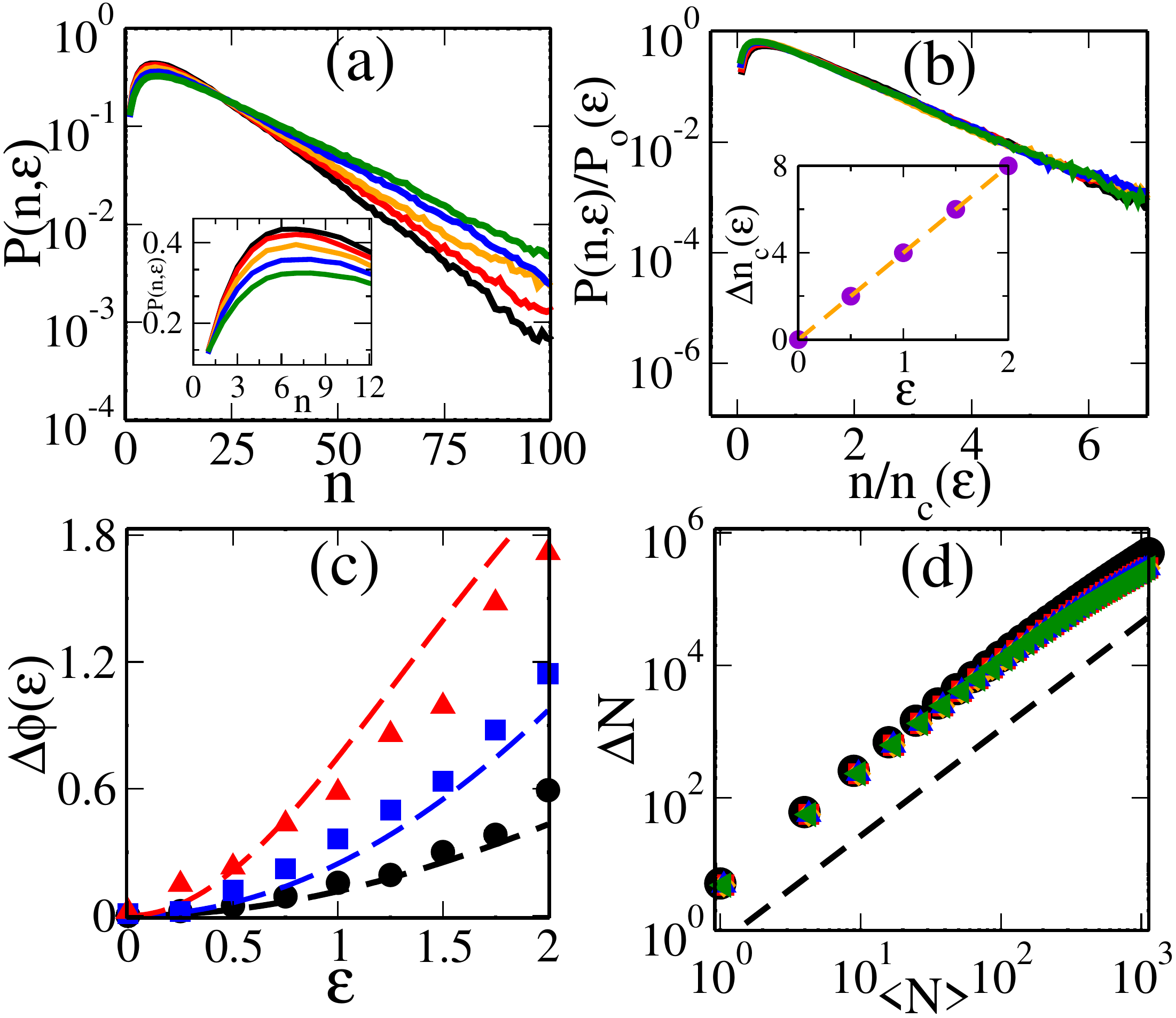}
	\caption{(color online) (a,b) 
	Plots of $P(n,\epsilon)$ vs. $n$ and
        $P(n,\epsilon)/P_{o}(\epsilon)$ vs. $n/n_c$ in semi-$log$ Y scale
	for different $\epsilon$, respectively. The black, red, orange, blue  and  green lines
        represent $\epsilon = 0.0, 0.5, 1.0, 1.5$, and $2.0$,
        respectively. Inset of (a) is zoomed near to the peak of the distribution of main plot. $N = 62500$. In the $inset$ of (b) 
	Plot of $\Delta n_c(\epsilon)= n_c(\epsilon)-n_c(0)$ vs. $\epsilon$.
	Voilet circles represent the data obtained from 
	the fitting function $exp(-n/n_c{(\epsilon)})$ and orange dashed line indicates 
	the linear variation.  
        (c) Variation of local density fluctuation $\Delta \Phi (\epsilon)$ with $\epsilon$.
	The filled black circles, blue squares and red triangles represent the numerical data 
	points for $\rho_0=0.5 ,1$ and $2$, respectively. Error bars are in the order of symbol sizes. $N = 62500$.
        The black ($\rho_0=0.5$), red ($\rho_0=1.0$) and blue ($\rho_0=2.0$) dashed line indicate
	the variation of $\Delta \Phi (\epsilon)$ obtained from the analytical
	calculations, as shown in Appendix.\ref{App:AppendixA} in Eq.(\eqref{q22}).
        (d) Plot of the global number fluctuation
	($\Delta \mathcal{N}$) vs. the mean number
	of particles $\langle \mathcal{N} \rangle$
        in $\log-\log$ scale. $N = 62500$. The dashed line represents
	slope $= 1.6$.}
\label{fig:fig4}
\end{figure}

As discussed in the previous paragraph, the disorder 
does not affect the usual long-range ordering in the 
system. Furthermore, we study the effect of the disorder 
on the  clustering of particles in the steady state. 
The snapshots of the system for three different strengths
of the disorder, $\epsilon =0 , 1$ and $2$ at different times 
are shown in Fig.\ref{fig:fig3} .
At late time, we note that the number of particles inside a unit sized 
cell increases for high disorder strength, as shown in Fig.\ref{fig:fig3} .
Hence, the  particles cluster more cohesively for high disorder strength $\epsilon$.
To further characterise the density clustering, we 
 calculate the probability distribution function (PDF) $P(n, \epsilon)$
 of the number of particles $(n)$ inside the
interaction radius for different  $\epsilon$. 
$P(n, \epsilon)$ for different $\epsilon$ decay with an 
exponential tail, $P(n, \epsilon) \sim P_o(\epsilon)\exp(-n/n_c(\epsilon))$ where $n_{c}$ is a constant and it is obtained from the exponential fitting, as shown
in Fig.\ref{fig:fig4}(a). The distribution flattens 
with the increasing strength of the disorder.
	Therefore, the particles
are having more number of neighbors inside its interaction radius, i.e.
more compact/dense clustering in the system. In the inset of Fig.\ref{fig:fig4}(a), the variation of P(n) with 'n' is shown. We note that the peak of the distribution decreases with the disorder strength. It further confirms that the probability of the small clusters is less for high disorder strengths.
In the  Fig.\ref{fig:fig4}(b), the scaling plot 
of $P(n, \epsilon)/P_o(\epsilon)$ \textit{vs.} $n/n_c(\epsilon)$
is shown for different $\epsilon$. 
We note that $n_{c}$ increases linearly with the disorder strength $\epsilon$,
as shown in the inset of Fig.\ref{fig:fig4}(b). It also suggests that the 
number of neighbors for each particle is increasing with $\epsilon$.
Therefore, the scaling behavior of the PDFs confirm that the 
clusters are statistically identical for 
different strengths $\epsilon$ of the disorder. 
To further understand the density clustering, 
we calculate the local density fluctuation, 
$\delta\phi(\epsilon)=\sqrt{\frac{1}{L^2}{\sum^{L^2}_{j=1}}{(\phi_j(\epsilon))^2}-(\frac{1}{L^2}{\sum^{L^2}_{j=1}}{\phi_j(\epsilon)})^2}$, for different $\epsilon$. 
To calculate $\delta \phi(\epsilon)$, we divide the full 
$L \times L$ system into
$L^{2}$ number of unit sized sub-cells. 
 $\phi_{j}(\epsilon)$ is 
the number of particles in the $j^{th}$ unit sized sub-cell and 
$\delta \phi(\epsilon)$ is the measure of the standard 
deviation in number of particles in a unit sized sub-cell of the system. 
Furthermore, we define the relative density phase separation by 
$\Delta \Phi (\epsilon) = \delta \phi(\epsilon)-\delta \phi(0)$, 
where $\delta \phi(0)$ is the local  density fluctuation for the clean system ($\epsilon=0$).

The plot of $\Delta \Phi(\epsilon)$ \textit{vs.} $\epsilon$ for three
different densities $\rho_0=0.5, 1.0$ and $2.0$
is shown in Fig.\ref{fig:fig4}(c). We note that
the density clustering increases with $\epsilon$ for all the densities.
We also calculate the magnitude of the density fluctuation 
using linearized hydrodynamic equations of motion 
for the coarse-grained density and orientation fields of the system. 
The dashed lines in Fig.\ref{fig:fig4}(c) is obtained from the 
linearized hydrodynamics in Eq.(\eqref{q22}). Since the linearized hydrodynamic 
works well in the mean field limit,  hence, the  data matches well for lower density and deviates 
for the higher densities. The details of the hydrodynamic calculation are 
given in Appendix \ref{App:AppendixA}. 
Therefore, the random-bond disorder which has a 
tendency to disturb the ordering in the corresponding 
equilibrium system \cite{MKumar, bishopprl},
enhances the density clustering in RBDPF. Hence, the disorder 
introduces more {\em cohesion} among the SPPs. \\
Furthermore, we calculate the global number fluctuation in different sub-systems, 
$\Delta \mathcal{N} =\sqrt{\langle {N}^2 \rangle -\langle{N} \rangle^2}$, 
$N$, $N^{2}$, and $\Delta\mathcal{N}$
represent the number of particle in a box of size $l$, the square of the number of particles in a
box of size $l$ and standard deviation, respectively. We varied $l$ from $1$ to
the $1/4$ of the system size. $\langle \rangle$ represents
the average over many snapshots and many ensembles.
We show the plot of $\Delta \mathcal{N}$ \textit{vs.} the mean number of particles 
in the sub-system $\langle \mathcal{N} \rangle$ for different $\epsilon$ in Fig.\ref{fig:fig4} (d). 
Although the disorder enhances the local density clustering, $\Delta \mathcal{N}$
remains unaffected in the presence of the disorder and the system shows the usual Giant number 
fluctuation for all $\epsilon$.
Also, we note that $\Delta \mathcal{N}  \simeq \langle \mathcal{N} \rangle ^{1.6}$, and it matches well with the previous 
studies of polar self-propelled particles interact through the Vicsek type 
interaction \cite{Chateprl2004,Chatepre2008, Biplabtopo}.
\begin{figure}[t]
\centering
\includegraphics[width=1.0\linewidth]{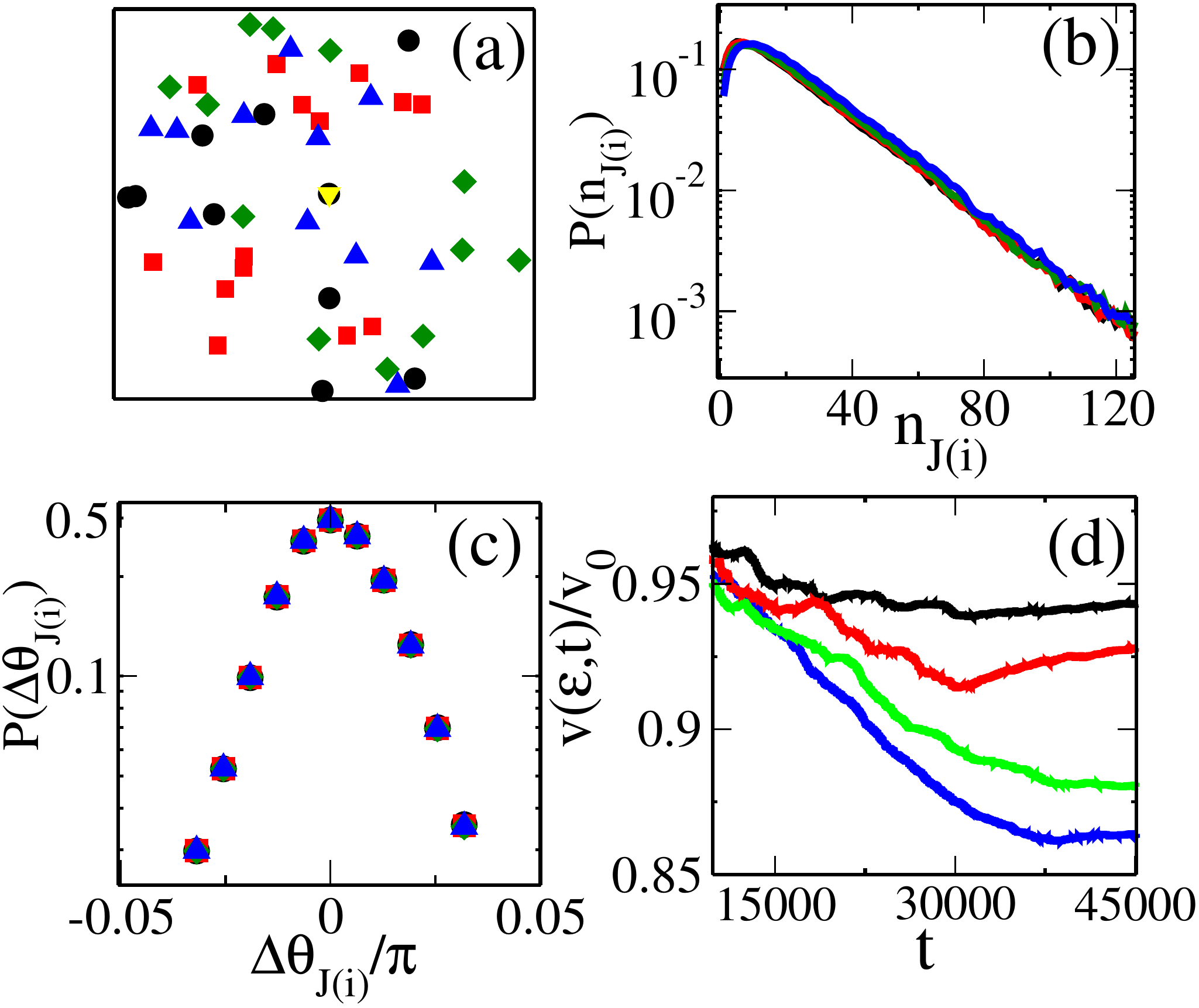}
	\caption{(color online) (a) Snapshot of the neighbour
        particles within the interaction radius ($R=1.0$) of a tagged particle 
	in the steady state for disorder strength $\epsilon=2.0$. $N=10000$.
        The filled black circles, red squares, green diamonds, and
        blue triangles denote $0 \leq J < 0.5$, $0.5 \leq J < 1.0$,
        $1.0 \leq J < 1.5$, and $1.5 \leq J \leq 2.0$, respectively. 
	At the centre of the box, yellow triangle indicates the tagged particle.
         (b,c) Plots of $P(n_{J(i)})$ and $P(\Delta\theta_{J(i)})$ 
	 distribution for different $J(i)$ ranges, respectively.
	$N =  62500$. Color lines and symbols in (b,c) indicate same things as in (a). 
	(d) Plot of normalise effective transport speed  $v(\epsilon,t)/v_0$ vs. $t$.
	Black, red, green and blue lines
        represent $\epsilon=0.0$, 1.0, 1.5 and 2 respectively.
        $N = 10000$.}
\label{fig:fig5}
\end{figure}
\subsubsection*{Distribution of particles in flock}
In the previous section, we note that the random-bond disorder 
introduces more cohesion among the SPPs.
To understand this mechanism of cohesion for higher disorder, 
we analyse a cluster and study the distribution of 
particles inside it, as shown in Fig.\ref{fig:fig5} .
The snapshot of particles' position inside the interaction 
radius of a tagged particle is shown 
in Fig.\ref{fig:fig5}(a).  
We divide the full range 
of $J \in [0,2]$ (for maximum disorder $\epsilon=2$) in 
four parts $J(1) \in [0:0.5]$, $J(2) \in [0.5,1.0]$, 
$J(3) \in [1.0:1.5]$ and $J(4) \in [1.5:2.0]$ and they are shown by different colors.
The snapshot shows that the particles of different interaction 
strengths are distributed homogeneously inside an interaction radius of a given particle. 
Furthermore, we calculate the probability distribution function (PDF) $P(n_{J(i)})$ 
of the particles $n_{J(i)}$ of the four different ranges of $J(i)$, where $i=1,2,3,4$, as 
shown in Fig\ref{fig:fig5}(b). 
We note that $P(n_{J(i)})$  
for each range of $J(i)$ are nearly identical and it confirms
that particles are distributed homogeneously in the system.
We also 
plot the particles orientation distribution 
$P(\Delta\theta_{J(i)})$ for the four different ranges of $J(i)$. 
The orientation distribution of the particles of different ranges,
$P(\Delta\theta_{J(i)})$, remains unchanged, 
as shown in the Fig.\ref{fig:fig5}(c). 
Hence, the clusters are a homogeneous network of particles of different interaction strengths for the RBDPF.
 Therefore, a moving particle always experiences a 
random network of interaction strengths during its motion. The resulting orientation
due to the random strength of neighbours results in the weaker alignment, hence, the system has less ordering. 
Furthermore,  
we calculate the effective transport speed $v(\epsilon,t)$ 
of the particles for different strengths of the disorder.
The mean displacement of the particles is calculated by 
taking the square root of their mean square displacement, $\Delta{r}(t)=\sqrt{\langle \Delta^{2} r(t)\rangle}$ where $\Delta^2{r(t)}=\sum^{N}_i\big(\langle r{^2}_i(t)\rangle -\langle r_i(t)\rangle^2\big)$. 
Moreover, the transport speed of the particles is defined as $v(\epsilon,t)=\frac{\Delta{r(t)}}{t}$. 
In Fig.\ref{fig:fig5}(d), we show the variation of the normalized effective transport speed $\frac{v(\epsilon,t)}{v_0}$ for four values of disorder strengths $\epsilon$ (= 0, 1, 1.5 and 2). We note that $\frac{v(\epsilon,t)}{v_0}$ decreases with increasing strength of the disorder. Hence, we claim that due to random nature of different interaction strength, dynamics of the particle become slow for high disorder strength. It further leads to strong clustering on increasing the strength of disorder.

\subsection{Dynamical Behaviour\label{Dynamical Behaviour}}
\subsubsection*{ Ordering kinetics to the steady state}
\begin{figure}[t]
\centering
\includegraphics[width=1.0\linewidth]{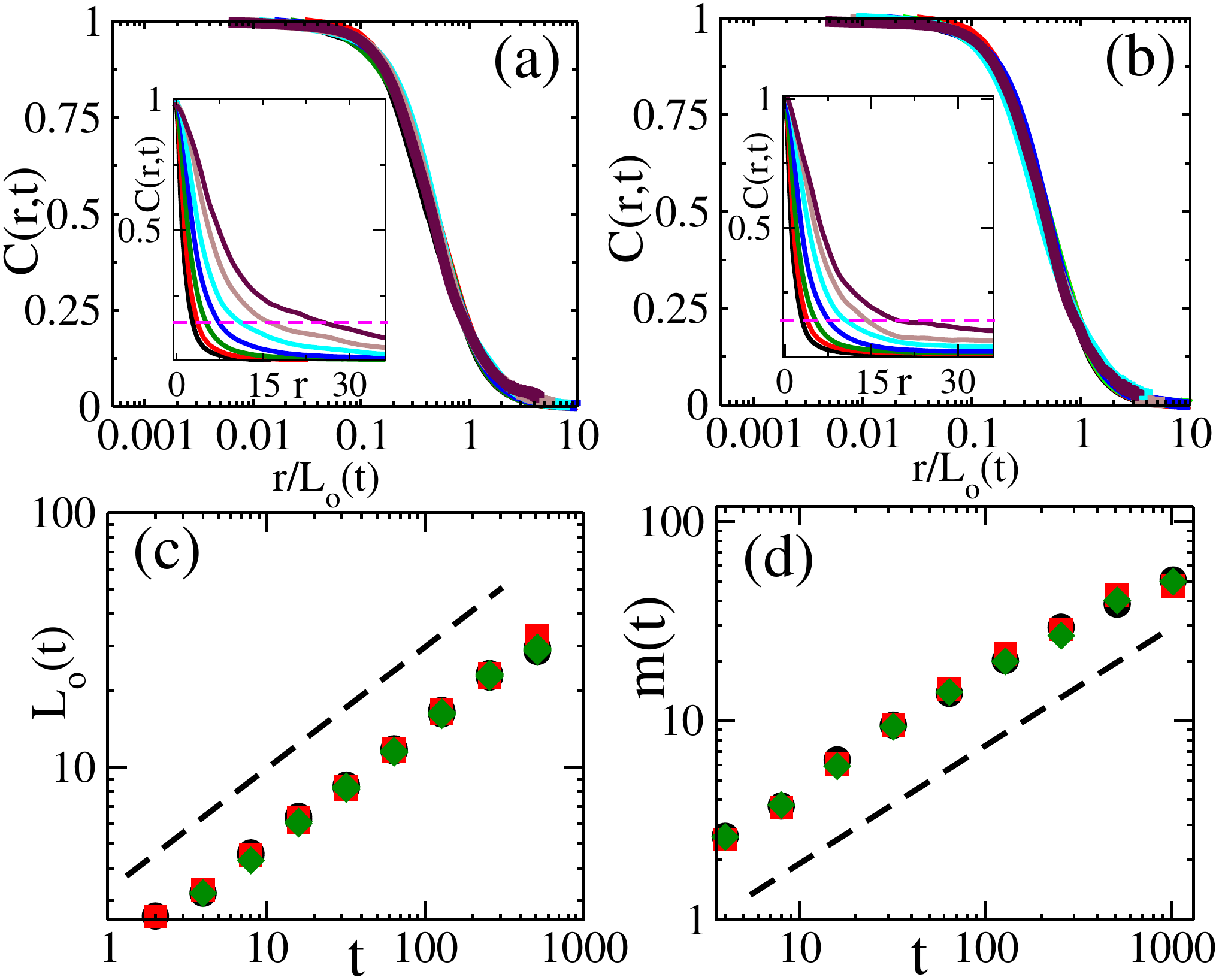}
	\caption{(color online) (a,b) Plots $C(r,t)$ vs. 
	$r/L_{o}$ for $\epsilon=0$ and $2$, respectively in semi-$\log$ Y-scale. 
	 Black, red, green, blue, cyan, brown and voilet lines represents 
	 time $t= 2, 4, 16, 32, 64, 128$ and $256$, respectively. 
	 In the $inset$ of (a) and (b), the variation $C(r,t)$ with 
	 $r$ is shown at different time. Different colored lines represent same thing as the main figure. Dashed line with color magenta is drown parallel to x-axis and intersect y-axis at 0.17 (crossing point) .
	 $N=262144$
	(c) Plot of $L_{o}$(t) with time $t$ in $\log-\log$ scale.
        The dashed line represents
        the slope $0.5$. $N = 262144$.
        (d) Plot of mass of the largest cluster $m(t)$
        with time $t$ in $\log-\log$ scale. $N = 40000$.
        The dashed line represents the slope $ 0.5$.
        The  filled black circles, red squares,
	and green diamonds represent $\epsilon=0.0, 1.0$ and $2.0$, respectively.}
\label{fig:fig6}
\end{figure}

In previous sections, 
we have discussed the steady-state properties of the 
ordered state. In this section, we discuss the effects of the 
random-bond disorder on the ordering kinetics when the system
is quenched from a random disordered
state to an ordered steady state. Kinetics of the orientation ordering 
is characterised by calculating the  
two-point orientation correlation function 
$C(r,t) = \big\langle \frac{\sum_{ij}{\bf {n}}_i({\bf {r}}_0,t)\cdot {\bf {n}}_j({\bf {r}}+{\bf {r}}_0,t)}{N(N-1)}\big\rangle - \frac{\sum_{ij}<{\bf {n}}_i({\bf {r}}_0, t)> <{\bf {n}}_j({\bf {r}} + {\bf {r}}_0, t)>}{N(N-1)}$,
where second term on the right hand side is zero.
$\langle $..$ \rangle$ represents average over many reference points $r_0$s' and $10$ independent realizations. 
We note that $C(r,t)$ grows with time for 
all disorder strengths $\epsilon$ as shown in $inset$ of Fig.\ref{fig:fig6}(a),(b). 
In the main plot of Fig.\ref{fig:fig6} (a) and (b), we find nice scaling with respect to the reduced 
length $r/L_o(t)$, where 
$L_{o}(t)$ is the characteristic domain size and it is
obtained from the first 0.17 crossing of the correlation function $C(r,t)$. 
The plot of $L_{o}(t)$ \textit{vs.} 
time $t$ for the clean system $\epsilon=0$ and for the RBDPF
 ($\epsilon=1,2$) are shown 
in Fig.\ref{fig:fig6}(c). We note that the 
disorder has no effect on the kinetics of growing domains.
Moreover, the size of domains varies as, 
$L_{o}(t) \simeq t^{1/z_{o}}$ with $z_{o} \sim 2$ 
for all disorder strengths.
We also calculate the kinetics of the growth of the density  cluster. Density growth is measured  by the 
 mass of the largest cluster $m(t)$. The mass of the largest
cluster $m(t)$ is calculated using 
the cluster counting algorithm \cite{Beat2017}. 
The plot of $m(t)$ {\em vs.} time $t$ for the clean system $\epsilon=0$ and the RBDPF, $\epsilon=1,2$, are shown in Fig. \ref{fig:fig6}(d).
For all cases,
$m(t)$ grows with time as $t^{\alpha}$  with $\alpha \sim 0.5$.
Hence, the length of the density cluster $L_{\rho}(t) \simeq \sqrt{m(t)}  \sim t^{1/z_{\rho}}$ and 
$z_{\rho} \sim 4$,
which is similar to the  asymptotic growth exponent 
for the conserved field in the active model B 
\cite{Cates2014, SudiptamodelB}.
\section{Discussion}
\label{Discussion}
We introduced a minimal model of a collection of self-propelled particles with the random-bond disorder. 
Each particle has a different ability (interaction strength) to influence its neighbours.
The varying interaction strength is obtained
from a uniform distribution and it can be varied from $[1-\epsilon/2: 1+\epsilon/2]$, where $\epsilon$ is
the disorder strength. For $\epsilon=0$, the model reduces to the
uniform interaction strength model or the Vicsek-like model \cite{VicsekT}. The equilibrium analogue of the present  
model is the random-bond XY model.
We studied the characteristics of the ordered steady state
for different strengths of the disorder. The random-bond disorder does not affect the
usual LRO present in a clean polar flock. To our surprise, the random-bond disorder leads to a more cohesive flock, 
hence, more
inhomogeneous or dense clusters. This phenomenon is due to the
slower dynamics of a particle moving in a random 
network of different interaction strengths.
Although the disorder affects the local density inhomogeneity, 
the global density fluctuation remains unaffected and the system shows the
 usual giant number fluctuation (GNF). \\
Furthermore, we also studied the effects of the random-bond disorder on the ordering kinetics of the orientation 
and the density fields.
We note that the orientation field in a polar flock with uniform interaction 
coarsens with time with a growth exponent $z_{o} \sim 2$ whereas the growth exponent for the
density field $z_{\rho} \sim 4$. Moreover, the coarsening for both the fields remain unaffected 
in the presence of random-bond disorder as opposed to what is observed
in the corresponding equilibrium model \cite{MKumar, bishopprl}.\\
Hence, our study introduces the effect of the random-bond disorder in a polar flock and 
shows many interesting features that are in
general not present in the corresponding equilibrium system with random-bond disorder \cite{MKumar, bishopprl}. 
Our study provides a new direction to 
understand the effects of intrinsic inhomogeneity in many natural active systems.
\section{Acknowledgement \label{Acknowledgement}}
J. P. Singh thanks computational facility at IIT(BHU), Varanasi. 
S. Pattanayak thanks TUE computational facility at S.N.B.N.C.B.S. 
S. Mishra thanks Department of Science and Technology, 
Science and Engineering Research Board (India), project no. ECR/2017/000659 for
partial financial support. 
\appendix 
\section{Linearised study of hydrodynamic equations of motion}\label{App:AppendixA}
We define local density of the particles as, 
\begin{equation}
	\rho({\bf r},t)=\sum_{i=1}^{N}\delta({\bf r}-{\bf R}_i(t))
    \label{q1}
\end{equation}
where, ${\bf R}_i$ and $N$ are the position vector of the $i^{th}$ particle and total number of particles respectively.  
Similarly we define the local polarisation density field as 
\begin{equation}
	{\bf P}({\bf r}, t) \rho={{\sum}^N_{i=1}{\bf n}_i(t)\delta({\bf r}-{\bf R}_{i}(t))}.
    \label{q2}
\end{equation}
We can write the coupled hydrodynamic equations of motion for density and polarization fields
as obtained in the study  \cite{TonerTu1998,Ebertin,Thle}.\\
\begin{widetext}
\begin{equation}
\partial_t\rho=-v_0\nabla\cdot({\bf P}\rho)
\label{q3}
\end{equation}
\begin{eqnarray}
    \partial_t{\bf P}& =&\bigg (\alpha_1(\rho, \epsilon)-\alpha_2 {\bf P}\cdot{\bf P}\bigg ){{\bf P}} 
	- \frac{v_1}{2\rho_0} \nabla ( \rho)  +\lambda({\bf P}\cdot\nabla){\bf P}+D_{p}\nabla^2{\bf{\rho	}}+{\bf f}_{\bf p}({\bf r}, t)
    \label{q4}
\end{eqnarray}	
For the random-bond disorder model, we introduce an additional general $\epsilon$ dependence term to alignment
	parameter $\alpha_{1}$ in the Eq.(\eqref{q4}). $\alpha_{1}$ is considered as a constant in the 
	study by Toner \textit{et al.} \cite{TonerTu1998} whereas have density and noise dependence in \cite{Ebertin,Thle}. 
	Here we show in steps, how we get the specific  depdence of $\alpha_1$ on $\epsilon$. 
	Starting from the position and orientation update (without repulsion) as given in Eq.(\eqref{q1}) and (\eqref{q2}), 
	we can write 
	\begin{equation}
		\partial_t({\bf {P}} \rho) = \frac{\sum_i\big[ {\bf {n}}_i(t+\Delta t) \delta({\bf {r}} - {\bf {R}}_i(t+\Delta t))- {\bf {n}}_i(t) \delta({\bf {r}} - {\bf {R}}_i(t))\big]}{\Delta t}
\label{q5}
	\end{equation}
	The above discretisation of time derivative are written by approximating all the relevant time scales are much larger than the unit time step of updation.  
	${\bf n}_i(t+\Delta t)$ and ${\bf R}_i(t+\Delta t)$ is obtained from the two upadtes of orientation and poistion as given in Eq.(\eqref{eq1}) and (\eqref{eq2}) and we have taken $\Delta t=1.0$ in the simulation.
	After substitution of the orientation and position at time $t+\Delta t$, we find
	\begin{equation}
		\partial_t({\bf P} \rho) = \sum_i\frac{[\sum_j J_j {\bf n}_j + N_i \phi_i \eta][\delta({\bf r}-{\bf R}_i(t)-v_0 {\bf n}_i(t))]}{\omega_i}-\sum_i[{\bf n}_i(t) \delta({\bf r}-{\bf R}_i(t))]
		 \label{q6}
	\end{equation}
	where $\omega_i = ||\sum_j {\bf n}_j(t) + N_i \phi_i \eta||$, where $|| .. ||$ means norm of the vector inside. After using the distribution $P(J_j) $ from 
	the uniform distribution of $J_j$  from $[1-\epsilon/2: 1+\epsilon/2]$ and use this in the summation inside the interaction radius and replacing it 
	the summation by the integral (since the distribution of $J$ is continuous), we find that the first term (linear term in ${\bf P}$) in Eq. (\eqref{q4}) will be 
	of the form $\alpha_1(\rho,\epsilon)= \alpha_0\bigg(\rho_0(\frac{1+\epsilon^2/12}{1-\epsilon^2/84})-\eta^2\bigg)$, 
	where $\alpha_0$ is a constant. Another derivation of the hydrodynamic equations from the  microscopic update equation 
	is similar to  as given  in \cite{ebertin, SudiptaJPCOM}. For simplicity, 
	we assume all other terms are independent of the disorder.\\ 
        Eq.(\eqref{q3}) represents continuity equation for the conserved density field $\rho$
	with a flux controlled by, $-v_0\nabla\cdot({\bf P}\rho)$ describes
	convection due to self-propulsion speed $v_0{\bf P}$.
	In Eq.(\eqref{q4}), the first term on right hand side represents a mean field 
	transition from an isotropic state (${\bf P}=0$) to a broken symmetry state
	${\bf P}  = \sqrt{\frac{\alpha_1 (\rho_0, \epsilon)}{\alpha_2}}
	{\widehat{\bf x}}$ (the direction of broken symmetry is chosen along $x-$axis). 
	The second and third term indicate hydrostatic pressure due to density
	gradient and convection in the model, respectively. Both $\lambda$ and $v_1$ depends on 
	self-propelled speed of the particle \cite{Ebertin}. Here in present study we assume it $\sim v_0$. The fourth  term represents
	diffusion in the polarisation field. The last term is noise in the system ${\bf f}_{\bf p}({\bf r}, t) = (f_{p_x}({\bf r}, t), f_{p_y}({\bf r}, t)) $is white Gaussian white noise with mean zero and variance $\Delta_{p}$.
	We perturb the system about the homogeneous steady state solution of Eqs. (\eqref{q3}) and (\eqref{q4}) 
	and write $\rho({\bf r},t)=\rho_{0} + \delta\rho$
        and ${\bf P}({\bf r},t)=(p_0+\delta{p_x}(t)){\widehat{\bf x}}+(\delta{p}_{y}(t)){\widehat{\bf y}}$, 
	where $p_0(t)= \sqrt{\frac{\alpha_1(\rho_0, \epsilon)}{\alpha_2}}$. 
	We write the linearised hydrodynamic equations for small perturbations 
	in three fields $\delta \rho(t)$, $\delta p_{x}(t)$ and $\delta p_{y}(t)$ as, 

	\begin{eqnarray}
		\partial_t \delta{p}_x = \bigg({\alpha}^{'}_1(p_0)p_0-\frac{v_1}{2\rho_0}\partial_{x}\bigg)\delta\rho-2 \alpha_1(\rho_0,\epsilon)\delta p_{x} 
\label{q7}
\end{eqnarray}
\begin{equation}
	\partial_t\delta{p}_{y}(t)=\lambda{p_0\partial_x\delta{p}_y}+D_{p}\nabla^2\delta{p_y}-\frac{v_1}{2\rho_0}\partial_y\delta\rho + f_{p_y}({\bf r}, t),
\label{q8}
\end{equation}
\begin{equation}
\partial_{t}\delta\rho(t)=-v_0(\partial_x(p_0+\delta{p_x})(\rho_0+\delta\rho)+\partial{y}\delta{p_y}(\rho_0+\delta\rho))
\label{q9}
\end{equation}
	where $\delta p_{x}$ and $\delta p_{y}$ are in the directions of broken symmetry and perpendicular to it, respectively and 
	$\alpha_1' = \frac{\partial\alpha_1(\rho)}{\partial\rho}\vert_{\rho_0}=\alpha_0\frac{(1+\epsilon^2/12)}{(1-\epsilon^2/84)}$.
In writing Eq.(\eqref{q7}), we assumed that fluctuations in the longitudinal direction is long range and higher order derivatives are neglegible.  In the steady state,  using Eq.(\eqref{q7})
	we can solve for $\delta p_x$
\begin{equation}
	\delta{p}_x=\frac{(\alpha'_{1}p_0-\frac{v_1}{2\rho_0}\partial_{x})\delta\rho}{2\alpha_1(\rho_0, \epsilon)}.
\label{q10}
\end{equation}
	We substitute $\delta p_{x}$ from Eq.(\eqref{q10}) in Eqs. (\eqref{q8}) and (\eqref{q9}) and write 
	effective dynamical equations for $\delta p_{y}$ and $\delta \rho$ as,
\begin{equation}
\partial_{t}\delta{p_y}=\lambda p_0\partial_{x}\delta{p_y}+D_p\nabla^2p_y-\frac{v_1}{2\rho_0}\partial_y\delta\rho
\label{q11}
\end{equation}
\begin{equation}
	\partial_{t} \delta\rho=v_0p_0V_x\partial_x\delta\rho+D_{\rho}{\partial_x}^2\delta\rho-v_0\rho_0\partial_y\delta{p}_y
\label{q12}
\end{equation}
where, $V_x=(\frac{\rho_{0}\alpha_{1}^{'}}{2\alpha_{1}}+1)$, $D_{\rho}=\frac{v_0v_1}{4\alpha_{1}}$ 
and $\alpha_1 = \alpha_1(\rho_0, \epsilon)$. 
Furthermore, we take the Fourier transform of Eq.(\eqref{q11}) and (\eqref{q12}) using $Y({\bf r}, t) = \int{d {\bf k} \exp(-i( {\bf k} \cdot {\bf r} + \omega t)) Y({\bf k}, \omega)}$ and write different terms in matrix notation, 
\begin{equation}
M\begin{bmatrix}
\delta\rho \\{\delta p_y}
\end{bmatrix}
=
\begin{bmatrix}
	0 \\ f_{p_y}
\end{bmatrix}
\label{q13}
\end{equation}
 where, the coefficient matrix $M$ can be written as, 
\begin{equation}
M=\begin{bmatrix}
(-i\omega+iq_xv_0p_0V_x-D_{\rho}{q_x}^2)&(-v_0\rho_{0}iq_y\delta p_y)\\
(\frac{iv_1}{2\rho_0}q_y\delta\rho)&(-i\omega-\lambda p_0iq_x-D_p{q^2})\\
\end{bmatrix}
\label{q14}.
\end{equation}
The Eq.(\eqref{q14}) gives the two modes from the linearised 
hydrodynamics calculations,
\begin{equation}
\omega_\pm=C_\pm(\theta)q-i\Gamma_L[\frac{V_\pm(\theta)}{2C_2(\theta)}]-i\Gamma_\rho[\frac{V_\pm(\theta)}{2C_2(\theta)}]
\label{q15},
\end{equation}
where, $C_\pm(\theta)=\frac{\gamma+v_0V_x }{2}{\cos\theta}\pm C_2(\theta)$, $C_2(\theta)=\sqrt{\frac{(\gamma-v_0V_x)^2cos^2\theta}{4}+ \rho_0 v_1 sin^2\theta}$, $\gamma=-\lambda v_0$, $\Gamma_\rho(q)=D_{\rho}q^2_x$, $\Gamma_L(q)=D_p q^2$ and $V_\pm(\theta)=C_2(\theta)\pm\frac{\gamma-v_0V_x }{2}{\cos\theta} $. $\theta$ is the angle between flock direction and propagation vector ${\bf q}$, 
and $\Gamma_\rho(q)$ and $\Gamma_L(q)$ are the wave vectors dependent on damping. Using Eq.(\eqref{q13}) we get
\begin{equation}
\begin{bmatrix}
\delta\rho\\{\delta p_y}
\end{bmatrix}
=
M^{-1}\begin{bmatrix}
	0 \\ f_{p_y}
\end{bmatrix}
\label{q16}
\end{equation}
Therefore, solution for the fluctuations in $\rho$, 
$\delta\rho(q,\omega)=G_{\rho p}(q,\omega) f_{p_y}({\bf q},\omega)$, where the propagator $G_{\rho p}(q,\omega)$ can be written as, 
\begin{equation}
	G_{\rho p}(q,\omega)=\frac{v_0 \rho_ i q_y }{(\omega-C_+(\theta)q)(\omega-C_-(\theta)q)+{[i\omega(\Gamma_\rho(q)+\Gamma_L(q)) - iqcos\theta(\gamma\Gamma_\rho(q)+v_0V_x\Gamma_L(q))]}}
\label{q17}.
\end{equation}
Furthermore, the two-point density-density correlation function, $C_{\rho\rho} = \langle |\delta\rho(q,\omega)|^2 \rangle$, can be written as,
\begin{equation}
	C_{\rho\rho}=\frac{v_0^2 \rho_0^2 q_y^2 \Delta_p }{(\omega-C_+(\theta)q)^2(\omega-C_-(\theta)q)^2+[\omega(\Gamma_\rho(q)+\Gamma_L(q))-q cos\theta(\gamma\Gamma_\rho(q)+v_0V_x\Gamma_L(q))]^2}
\label{q18}.
\end{equation}
Moreover, the density fluctuation $\langle |\delta \rho({\bf q}, \omega)|\rangle$ can be obtained as,
\begin{equation}
\sqrt{C_{\rho\rho}}=\bigg[\frac{v_0\rho_0\sin\theta\sqrt{\Delta_p}}{C_+(\theta)(\Gamma_\rho+\Gamma_L)-\cos\theta(\gamma\Gamma_\rho+v_0V_x\Gamma_L)}\bigg]+\bigg[\frac{v_0\rho_0\sin\theta\sqrt{\Delta_p}}{C_-(\theta)(\Gamma_\rho+\Gamma_L)-\cos\theta(\gamma\Gamma_\rho+v_0V_x\Gamma_L)}\bigg]	
\label{19}.
\end{equation}
We can write Eq.(\eqref{19}) in a simple form for fluctuation in $\theta=\pi/4$ and for finite $q \sim 1$,
\begin{equation}
	\langle |\delta\rho| \rangle =\sqrt{\Delta_{p}}[\frac{(C_+(\theta) D_{\rho}-A)+(C_-(\theta) D_{\rho}-A)}{(C_+(\theta)D_{\rho}-A)(C_-(\theta)D_{\rho}-A)}]
\label{q21}.
\end{equation}
Substituting $A=\gamma D_{\rho}+v_0V_x\Gamma_L$, $C_++C_-=\frac{\gamma+v_0V_x}{2}$,  $C_+C_-=\gamma v_0V_x$ in Eq.(\eqref{q21}) and further simplification gives,
\begin{equation}
	\langle \vert{\delta\rho}\vert \rangle= \frac{\rho_0 \sqrt{\Delta_p} \alpha_0\bigg(\rho_0\frac{1+\epsilon^2/12}{1-\epsilon^2/84}-\eta^2\bigg)}{\gamma{v_1}+\rho_0\alpha_0(\frac{1+\epsilon^2/12}{1-\epsilon^2/84})+1}
\label{q22}.
\end{equation}
Substituting value of $\epsilon = [0.0, 2.0]$,  $v_0=0.5$, $\eta=0.2$ and  $\Delta_p = 0.1$ and using $\alpha_0$ as fitting parameter to match the curve for smallest
$\rho_0=0.5$  we found $\alpha_0=8$, curve matches well.  Then changed   value of $\rho_0=1.0$ and $2.0$ for other two curves. The  $\gamma$ and  $v_1$ $\sim v_0$. The plot of
 $\langle \vert{\delta\rho}\vert \rangle$ for three $\rho_0$ is shown by the dashed line in Fig.\ref{fig:fig4}(c).

\end{widetext}


\begin{thebibliography}{9}

\bibitem{Nedelec1997} F. N$\acute{e}$d$\acute{e}$lec, Ph.D. thesis, Universit$\acute{e}$ Paris {\bf 11}, 1998;
F. N$\acute{e}$d$\acute{e}$lec, T. Surrey, A. C. Maggs, and S. Leibler, Nature (London) {\bf 389}, 305 (1997).

\bibitem{Yokota1986} H. Yokota (private communication); Y. Harada,
A. Noguchi, A. Kishino, and T. Yanagida, Nature
(London) {\bf 326}, 805 (1987); Y. Toyoshima et al., Nature
(London) {\bf 328}, 536 (1987); S. J. Kron and J. A. Spudich,
Proc. Natl. Acad. Sci. U.S.A. {\bf 83}, 6272 (1986).

\bibitem{Garcia} S. Garcia, E. Hannezo, J. Elgeti et. al. PNAS {\bf 112} (50) 15314-15319 (2015). 

\bibitem{Bonner1998} J. T. Bonner, Proc. Natl. Acad. Sci. U.S.A. {\bf 95}, 9355
(1998); M. T. Laub and W. F. Loomis, Mol. Biol. Cell {\bf 9},
3521 (1998).

\bibitem{Chen2019} D. Chen , Y. Wang, G. Wu, M. Kang, Y. Sun, and W. Yu, Chaos {\bf 29}, 113118 (2019).

\bibitem{Parrish1997} Three Dimensional Animals Groups, edited by J. K.
Parrish and W. M. Hamner (Cambridge University
Press, Cambridge, England, 1997).

\bibitem{Helbing2000} D. Helbing, I. Farkas, and T. Vicsek, Nature (London) {\bf 407},
487 (2000).

\bibitem{VicsekT} T. Vicsek, A. Czir$\acute{o}$k, E. Ben-Jacob, I. Cohen, and O. Shochet, Phys. Rev. Lett. {\bf 75}, 1226 (1995).
\bibitem{Lucas}Lucas Barberis Phys. Rev. E {\bf 98}, 032607  (2018)
\bibitem{jpsingh} J.P Singh. Sameer Kumar and Shradha Mishra  (to be published).

\bibitem{Dgayer} D Geyer, D Martin, J Tailleur, D Bartolo. Physical Review X {\bf 9 }(3), 031043, (2019).
\bibitem{Petitjean}N Sepúlveda, L Petitjean, O Cochet, E Grasland-Mongrain, P Silberzan et. al. PLoS Comput Biol {\bf 9} (3), e1002944, (2013).
\bibitem{Caprini} L Caprini, UMB Marconi, A Puglisi. Phys. Rev. Lett. {\bf 124} (7), 078001, (2020).

\bibitem{TonerTu1998} J. Toner and Y. Tu, Phys. Rev. E {\bf 58}, 4828 (1998).

\bibitem{Ebertin} E. Bertin et al., Phys. Rev. E {\bf 74}, 022101 (2006)
\bibitem{ebertin}E Bertin, H Chaté, F Ginelli, S Mishra, A Peshkov, S Ramaswamy
New J. Phys. {\bf 15} (8), 085032 
\bibitem{Thle}  T. Ihle, Phys. Rev. E {\bf 83}, 030901 (2011),

\bibitem{Chateprl2004} G. Gr$\acute{e}$goire and H. Chat$\acute{e}$, Phys. Rev. Lett. {\bf 92}, 025702 (2004).

\bibitem{Chatepre2008} H. Chat$\acute{e}$, F. Ginelli, Guillaume Gr$\acute{e}$goire, and F. Raynaud,
        Phys. Rev. E {\bf 77}, 046113 (2008).

\bibitem{SudiptaJPCOM} S. Pattanayak and S. Mishra,  J. Phys. Commun, {\bf 2}, 045007 (2018).

\bibitem{Ihlepre2014} M. Romensky, V. Lobaskin, and T. Ihle, Phys. Rev. E {\bf 90}, 063315 (2014).

\bibitem{Morin2017} A. Morin, N. Desreumaux, J. Caussin, and D. Bartolo, Nature Physics {\bf 13}, 63–67 (2017).

\bibitem{Chepizhko2013} O. Chepizhko, E. G. Altmann, and F. Peruani, Phys. Rev. Lett. {\bf 110}, 238101 (2013).

\bibitem{Yllanes2017} D. Yllanes, M. Leoni, and M. C. Marchetti, New J. Phys. {\bf 19}, 103026 (2017).

\bibitem{Quint2015} D. A. Quint and A. Gopinathan, Phys. Biol. {\bf 12}, 046008 (2015).

\bibitem{Sandor2017} C. S$\acute{a}$ndor, A. Lib$\acute{a}$l, C. Reichhardt, and C. J. Olson Reichhardt, Phys. Rev. E {\bf 95}, 032606 (2017).

\bibitem{Reichhardt2017} C. J. O. Reichhardt and C. Reichhardt, Nat. Phys. {\bf 13}, 10 (2017).

\bibitem{Rakesh2018} R. Das, M. Kumar, and S. Mishra, Phys. R. E. {\bf 98}, 060602(R) (2018).

\bibitem{Toner2018E} J. Toner, N. Guttenberg, and Y. Tu, Phys. Rev. E {\bf 98}, 062604 (2018).

\bibitem{Toner2018L} J. Toner, N. Guttenberg, and Y. Tu, Phys. Rev. Lett. {\bf 121}, 248002 (2018).

\bibitem {RDas2020} R. Das, M. Kumar, and S. Mishra, Phys. Rev. E {\bf 101}, 012607 (2020).

\bibitem{SudiptaIS} S. Pattanayak, J. P. Singh, M. Kumar, and S. Mishra, Phys. Rev. E {\bf 101}, 052602 (2020). 

\bibitem{Lin2018} Guo-yuan Wang, Fan-yu Wu, You-liang Si, Q. Zeng, and P. Lin, Procedia Engineering
                {\bf 211}, 699 (2018).

\bibitem{Dorso2011} G. A. Frank and C. O. Dorso,  Physica A (Amsterdam, Neth.) {\bf390}, 2135 (2011).

\bibitem{ZuriguelJSM} A. Garcimartín, D. R. Parisi, J. M. Pastor, C. Martín-G$\acute{o}$mez, and I. Zuriguel, J. Stat. Mech. {\bf 4}, 043402 (2016).

\bibitem{Zuriguel2016} I. Zuriguel, J. Olivares, J. M. Pastor, C. Mart$\acute{i}$n-G$\acute{o}$mez,
L. M. Ferrer, J. J. Ramos, and A. Garcimartín, Phys. Rev. E {\bf 94}, 032302 (2016).

\bibitem{Zuriguel2011} I. Zuriguel, A. Janda, A. Garcimartín, C. Lozano, R. Ar$\acute{e}$valo, and D. Maza
        Phys. Rev. Lett. {\bf 107}, 278001 (2011).

\bibitem{VLBerezinskii} V. L. Berezinskii, JETP {\bf 32}, 493 (1971).

\bibitem{JMKosterlitz} J. M. Kosterlitz and D. J. Thouless, J. Phys. C {\bf 6}, 1181 (1973).

\bibitem{KosterlitzJPhys} J. M. Kosterlitz, J. Phys. C {\bf 7}, 1046 (1974).

\bibitem{CRoland} S. Puri and C. Roland, Phys. Lett. A {\bf 151}, 500 (1990).

\bibitem{SPuriPhys} S. Puri, Phys. Lett. A {\bf 164}, 211 (1992).

\bibitem{PuriDChowdhury} S. Puri, D. Chowdhury, and N. Parekh, J. Phys. A {\bf 24}, L1087
(1991).

\bibitem{WilliamPNAS} W. Bialek, A. Cavagna, I. Giardina, T. Mora, E. Silvestri, M. Viale, and A. M. Walczak,
PNAS {\bf 109}, 4786 (2012).

\bibitem {Zh} Vik. S. Dotsenko and M. V. FeTgel$^{'}$man, Zh. Eksp. Teor. Fiz. {\bf 83}, 345 (1982).

\bibitem {MKumar} M. Kumar, S. Chatterjee, R. Paul, and S. Puri,
Phys. Rev. E {\bf 96}, 042127 (2017).

\bibitem{bishopprl} D. J. Bishop and J. D. Reppy, Phys. Rev. Lett. {\bf 40}, 1727(1978).

\bibitem{Bray1994} A. J. Bray, Advances in Physics {\bf 43}, 357 (1994).
	
\bibitem{Puri2009} S. Puri, and V. Wadhawan, Kinetics of Phase Transitions, CRC press (2009).

\bibitem{Cates2014} R. Wittkowski, A. Tiribocchi, J. Stenhammar, R. J. Allen, D. Marenduzzo, and M. E. Cates,
	Nature Communications {\bf 5}, 4351 (2014).

\bibitem{SudiptamodelB} S. Pattanayak, S. Mishra, and S. Puri (to be published).

\bibitem{Biplabtopo} B. Bhattacherjee, S. Mishra, S.S. Manna, Phys. Rev. E. {\bf 92}, 062134 (2015).

\bibitem{Beat2017} C. P. Beatrici, R. M. C. de Almeida, and L. G. Brunnet, Phys. Rev. E. {\bf 95}, 032402 (2017).


\end{thebibliography}
\end{document}